\documentclass[iop]{emulateapj}

\usepackage{hyperref}
\usepackage{natbib}

\shorttitle{HD 95086 \MakeLowercase{b}}
\shortauthors{Meshkat et al.}

\begin{document}

\title{Further evidence of the planetary nature of HD 95086 \MakeLowercase{b} from Gemini/NICI \textit{H}-band data}

\author{T. Meshkat$^1$, V. Bailey$^2$, J. Rameau$^3$,  M. Bonnefoy$^4$, A. Boccaletti$^5$, E.~E. Mamajek$^6$, 
M. Kenworthy$^1$, G. Chauvin$^3$, A.-M. Lagrange$^3$, K. Y. L. Su$^2$, and T. Currie$^7$}

\altaffiltext{1}{Sterrewacht Leiden, P.O. Box 9513, Niels Bohrweg 2, 2300 RA Leiden, The Netherlands}  
\altaffiltext{2}{Steward Observatory, Department of Astronomy, University of Arizona, 933 North Cherry Avenue, Tucson, AZ 85721-0065, USA}
\altaffiltext{3}{UJF-Grenoble 1 / CNRS-INSU, Institut de Plan\'{e}tologie et d'Astrophysique de Grenoble (IPAG) UMR 5274, Grenoble, F-38041, France}
\altaffiltext{4}{Max Planck Institute f\"{u}r Astronomy, K\"{o}nigsthul 17, D-69117 Heidelberg, Germany}
\altaffiltext{5}{LESIA, Observatoire de Paris, CNRS, University Pierre et Marie Curie Paris 6 and University Denis Diderot Paris 7, 5 place Jules Janssen, 
F-92195 Meudon, France}
\altaffiltext{6}{Department of Physics and Astronomy, University of Rochester, Rochester, NY 14627-0171, USA}
\altaffiltext{7}{Department of Astronomy and Astrophysics, University of Toronto, 50 St. George St., Toronto, Ontario, M5S 1A1, Canada}


\begin{abstract}
We present our analysis of the Gemini/NICI \textit{H}-band data of HD 95086, 
following the discovery of the planet HD 95086 b  in \textit{L'}. 
The \textit{H}-band data reach a contrast of 12.7 mag relative to the host star 
at 5$\sigma$ levels in the location of HD 95086 b, and no point source is found. 
Our non-detection and $H-L'$ color limit rules out the possibility that the object 
is a foreground L/T dwarf and that, if it is bound to HD95086, it is a genuine planetary mass object.
We estimate a new pre-main-sequence isochronal age for HD 95086 of 17$\pm$4 Myr, which is commensurate with
previous mean age estimates for the Lower Cen-Crux subgroup.
Adopting an age of 17 Myr, the color limit is inconsistent with the COND model, 
marginally consistent with the BT-SETTL model, and consistent with the DUSTY model.

\end{abstract}

\section{Introduction}

\citet{Rameau13b} reported the probable detection of a 4 to 5 Jupiter mass companion to HD 
95086 lying at a projected separation of 56 AU. The host is a young (17$\pm$4 Myr; see Section \ref{sec:age}) A8 type star.
The large infrared excess \citep{Chen12} suggests that the star is harboring 
a debris disk, as yet unresolved.

The discovery at the \textit{L'} band (3.8 $\mu m$), using the angular 
differential imaging technique (ADI; \citealt{Marois06}), was a 9$\sigma$ detection in the first epoch.
The second epoch suffered from poor observing conditions, resulting in a 3$\sigma$ re-detection of the planet.
The astrometric precision allowed for the rejection of a background object with nearly 3$\sigma$ probability.
\citet{Rameau13b} also carried out \textit{Ks} (2.18 $\mu$m) observations that 
also suffered from bad observing conditions. This resulted in low \textit{Ks} sensitivity and a 
non-detection. These observations were used to conclude that the detection 
at \textit{L'} was likely inconsistent with a background star. Such a non-detection rejected the background 
hypothesis and suggested a red object with \textit{Ks-L'}$>$1.2 mag.

With a predicted temperature of $T_{\text{eff}} \simeq$ 1000 K 
(compared to the 1600--1700 K of $\beta$ Pic b, and 800--1100 K for HR8799 bcde; \citealt{Bonnefoy13,Currie11,Skemer12}), a giant planet 
will have an atmosphere close to the L-T spectral transition, where the effect of reduced surface 
gravity (characteristic of young and low-mass companions) is expected to dramatically 
affect the balance between dust formation and settling in the atmosphere \citep{Stephens09}.
If the companion is real, it would be a benchmark for spectroscopic studies of low-mass 
giant planets since it would be one of the lightest directly imaged planets, \footnote{HR8799 b \citep{Marois08} might 
be as light as HD 95086 b but its low mass estimate was derived from dynamical instability 
rather than from atmospheric models which give higher mass, and Fomalhaut b might be a 
dwarf planet and only the reflecting dust is detected \citep{Kalas08,Currie12b}.} and key to testing 
predictions about dust settling in low-gravity atmospheres.
The methane bands seen in mature brown-dwarfs at a similar temperature as HD 95086 b might also be 
reduced (or lacking) due to the effect of reduced 
surface gravity, which in turn triggers non-equilibrium chemistry of CO/CH$_4$\ \citep{Hubeny07}.

Low-resolution spectroscopy is highly challenging with the present instrumentation given 
the contrast and faintness of HD 95086 b \citep{Vigan12b}. Broad band photometry 
is currently the only way to derive of the color of the companion and constrain its atmospheric properties.
We examine our \textit{H}-band NICI data\footnote{Taken as part of a survey of young stars with 
debris disks (PI: V. Bailey).} taken in 2012, the same epoch as the \textit{L'} discovery image.
We do not detect the companion in our \textit{H}-band data, allowing us to derive a lower limit to 
the \textit{H-L'} color and reject a foreground contamination hypothesis. 

In Sections \ref{sec:obs} and \ref{sec:image_processing}, we describe our observations and our derived upper 
\textit{H}-band limit. In Section \ref{sec:analysis}, we discuss the age as determined 
by isochrone fitting and we show that the 
very red color rejects the foreground contamination hypothesis.

\section{Observations}
\label{sec:obs}
\subsection{Data}

Observations of HD 95086 were taken on UT 2012 March 19, 26, and 30 using the NICI camera \citep{Toomey03}
on Gemini South in the \textit{H}-filter ($\lambda$ = 1.65$\mu$m and $\Delta\lambda$ = 0.29$\mu$m) 
as part of a survey looking for gas giant planets around young stars 
with IR excesses (PI: V. Bailey). The camera was configured with the broad \textit{H}-band filter 
and the 0\farcs32 coronagraphic mask. The integration times for individual science 
frames are 55.0 s (5 coadded readouts of 11.0 s each). 
The details of the observing log for these data is listed in 
Table 1. Five percent of the frames are rejected per night due to poor alignment of the central star under 
the mask. The instrument was configured in telescope pupil tracking mode to keep 
the point spread function (PSF) structure from the telescope optics fixed with respect to the orientation of 
the detector. As a consequence, the position angle (PA) of the sky on the detector underwent 
rotation of 19$^{\circ}$, 25$^{\circ}$, and 25$^{\circ}$ on each of the nights, respectively.

\begin{table}[ht]
\scriptsize
\caption{Observing Log for NICI \textit{H}-band 2012 Data} 
\centering 
\begin{tabular*}{\linewidth}{p{2.9cm} l l l}
\hline
   Date (UT) & 2012 Mar 19 & 2012 Mar 26 & 2012 Mar 30  \\ [0.5ex]
  \hline\hline
   Total integration time (s)  &  2695 & 3520 & 3520 \\
   Frames observed & 49 & 64 & 64 \\
   Frames rejected & 5 & 4 & 4 \\
   On-sky rotation ($^{\circ}$) & 18.24 & 25.13 & 25.19 \\
   Parallactic angle ($^{\circ}$) & 18.89/37.13 &  342.14/07.27 &  347.71/12.90 \\ 
   DIM seeing ($\arcsec$)& 0.6 &  0.6 &  0.6 \\  [1ex]
    \hline
   \end{tabular*}  
\label{table:data}
\end{table}

\subsection{NICI Data Reduction}

Calibration data consists of dark frames with the same integration time and flats from 
the nights of March 19 and 26. The flat field images were dark-subtracted, and all individual 
flat field images were normalized by their median and combined together with a clipped mean 
to form a final normalized flat field image. 

The dark frames were subtracted from the science frames and then the dark-subtracted 
science frames were divided by the normalized flat field frame. A bad pixel mask was 
constructed from pixels in the flat field frame that were greater than 1.2 or less than 
0.8 in value. These bad pixels were interpolated over using adjacent good values in the science frames. Hot, cold, and 
flaky pixels were found and interpolated over in individual science frames by their 
values being more than 1000 counts different from the median of a $3\times3$ box centered on that pixel. 
Finally, the NICI distortion correction was applied to the images in order to perform astrometry 
on the additional background sources in the field of view.

\subsection{Photometric Calibration}
Due to the radial transmission function of the apodizer \citep{Wahhaj11} and the level of adaptive optics (AO) correction 
on the central star, photometric calibration was performed using a background star 
visible 4\farcs5 away from the primary star in the images.
It was first detected by \citet{Kouwenhoven05} 
and confirmed as a background object by \citet{Rameau13b}. 
This background star was used as a photometric reference to determine the 
flux calibration and sensitivity in our \textit{H}-band data. 
The NICI coronagraphic transmission curve is not yet published, thus to 
determine the near-infrared photometry of the background star, we utilize 
archival photometry from \textit{Hubble Space Telescope} (\textit{HST}) NICMOS data taken in 2007 (program 11157, PI: 
Joseph Rhee). The calibrated mosaic files provided by the \textit{HST} archive are background-subtracted 
and normalized by exposure time. Two images, one per roll angle, are taken 
in each F110W and F160W; each of the four images has a total exposure time of 895 s. 
We roll-subtracted each pair of images, and used the aperture diameter and corresponding 
aperture correction recommended by the NICMOS data reduction handbook for NIC2. 
We derived the magnitudes in F110W and F160W of $13.63\pm0.05$ and $12.81\pm0.05$ for the background 
star. We used the transformations of \citet{Stephens00} to derive \textit{J} and \textit{H} magnitudes of 
$13.25\pm0.10$ and $12.81\pm0.10$, respectively.

\section{Image Processing}
\label{sec:image_processing}

The data were centroided in a smaller frame which ensured the background visual 
binary star is clearly visible at an angular separation of 4\farcs5 and a PA of 
330$^{\circ}$. The unsaturated visual binary star is crucial for photometry, as the 
host star was blocked by the coronagraph, and the transmission of the light 
from the star is not well constrained. 
\citet{Kouwenhoven05} estimated this star to have a \textit{Ks} magnitude of 12.67 mag. 
\citet{Rameau13b} found the background star to have a $Ks-L'$ color 
of $-0.3\pm0.2$ mag. Using our NICMOS-derived \textit{H}-band magnitude of the 
background star, we estimate a $H-Ks$ color of 0.14$\pm$0.10 mag for the background star.

To understand the impact of imaging processing on the detectability of 
faint sources, we processed the data with artificially added planets with 
three independent pipelines: one by \citet{Meshkat13b} based on principal component analysis (PCA; \citealt{Amara12}), 
one by \citet{Boccaletti13} using locally optimized combination of images (LOCI; \citealt{Lafreniere07b}), and one by \citet{Chauvin12} using ADI.
The three pipelines agree with each other, and here we only show the results using PCA. 

The PCA pipeline was run separately for the data from each night, with an inner radius of 0\farcs50 
corresponding to the edge of the coronagraphic mask. For each dataset, five principal components 
were used to create a model of the stellar coronagraphic image. This optimal number of 
principal components was determined by injecting artificial components at the same angular 
separation with a different PA from the expected planet. The final de-rotated datacubes from 
each night are averaged for a final image.

\begin{figure}
 \centering
   \epsscale{1.1}
  \plotone{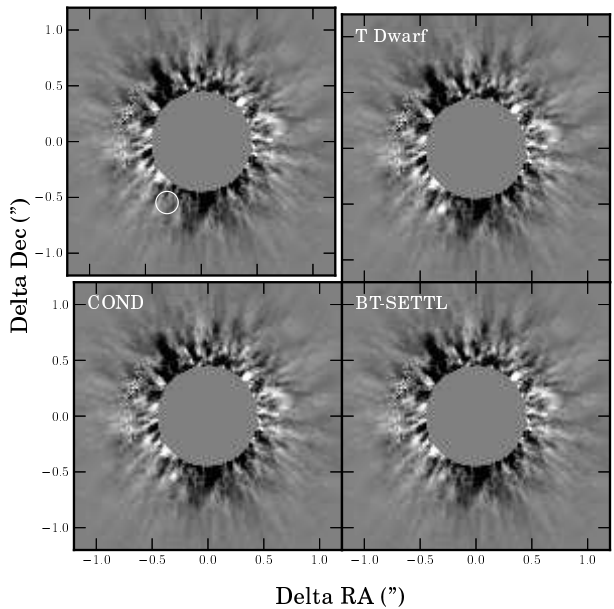}
 \caption{Upper left: the combined NICI \textit{H}-band image reduced with PCA with a white circle at the 
 expected location of the companion. The three other images have a fake companion injected 
 into the raw data at the location of the expected planet. The contrast of the artificial companions are 11.6 mag for the T dwarf, 
 12.0 mag for the COND model, and 12.5 mag for the BT-SETTL model.
}
 \label{hd95_fig}
\end{figure}

\autoref{hd95_fig} shows the PCA reduced image in the upper left corner. 
No point sources are seen at the expected position (see the white circle in \autoref{hd95_fig}).
The other images are 
with an artificial planet injected with the contrast and angular separation (0\farcs624) 
of the expected companion if it were a T dwarf (11.6 mag), COND model planet (12.0 mag), and BT-SETTL model planet (12.5 mag) 
(for details see Section \ref{sec:color_constraint}).

We injected artificial sources in all three nights of data combined to determine the 5$\sigma$ detection limit. 
The background star was used to generate the artificial planets by rescaling the flux to between 
10 and 14 mag fainter than the primary star, from 0\farcs55 to 0\farcs8  in 
0\farcs05 increments.

\begin{figure}
 \centering
  \epsscale{1.1}
 \plotone{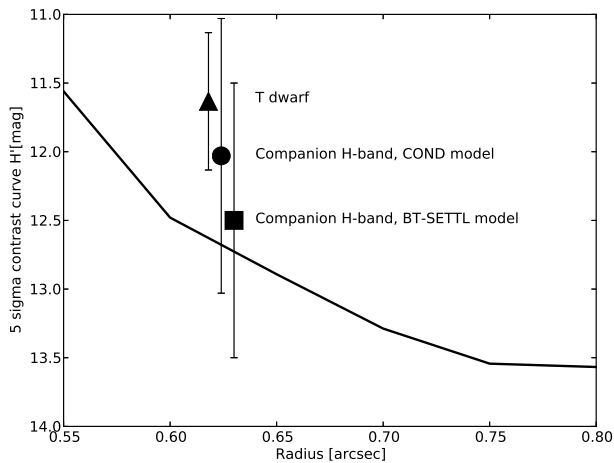}
 \caption{Contrast curve for \textit{H}-band NICI data on HD 95086 generated by injecting artificial planets in the three nights of data. The 
 circle, square, and triangle indicate the expected \textit{H}-band contrast for a 17 Myr COND model planet, BT-SETTL model planet, and 
 T dwarf, respectively. The square and triangle are slightly offset from the expected angular separation (0\farcs624) for clarity. }
 \label{contrast_curve}
\end{figure}

The signal-to-noise (S/N) of the resulting fake planet is determined by the following equation:

\[\left (\text{S/N}  \right )_{\text{planet}}= \frac{F_{\text{planet}}}{\sigma (r)\sqrt{\pi r_{\text{ap}}^{2}} }\]

where $F_{\text{planet}}$ is the sum of the flux in an aperture with a radius, $r_{\text{ap}}$, of 2 pixels, $\sigma$ is the 
root mean square of the pixels in a 340$^{\circ}$ arc at the same radius around the star 
(excluding the planet itself), and $r$ is the width of the arc (2 pixels). 
We can then construct the distribution of S/N at various source magnitudes, and determine the 
5$\sigma$ detection limit by interpolation at a given radius from the star. The 5$\sigma$ 
contrast curve is shown in \autoref{contrast_curve}.

To make sure the source brightness is not affected by the NICI coronagraph, we extrapolated 
the transmission measured in the CH$_4$\ filter \citep{Boccaletti13} assuming a linear 
relation with wavelength, and concluded that it should have nearly 100\% 
transmission at greater than 0\farcs6. Since the planet is detected at a separation of 0\farcs624, 
this is within the 100\% transmission regime.

\section{Analysis}
\label{sec:analysis}
\subsection{Stellar Parameters and Age}
\label{sec:age}

HD 95086 was kinematically selected as a member of Lower Cen-Crux
(LCC) based on its \textit{Hipparcos} astrometry by \citet{deZeeuw99}, and it
has been included in several later studies as a member
\citep[e.g.,][]{Rizzuto12, Chen12}. The {\it mean} age of the LCC
subgroup defined by the $>$1 $M_{\odot}$ stars is $\sim$16--17 Myr,
however, the inferred age spread of the group ($\sim$10 Myr) hints that
adoption of the mean subgroup age for a given star may be problematic
\citep{Mamajek02, Preibisch08, Pecaut12}. The revised \textit{Hipparcos}
astrometry for the star from \citet{vanLeeuwen07} differs negligibly
from the original value, and indeed \citet{Rizzuto12} again included
HD 95086 in a recent analysis, so we regard the kinematic membership
as secure.

The \textit{Hipparcos} catalog quotes photometry of $V =
7.36$ and $B-V = 0.230\pm0.004$. If the intrinsic colors of HD
95086 are similar to that of A8V dwarf stars, then their mean color is
$B-V = 0.25$ \citep{Pecaut13}, with an rms spread of $\pm$0.04
mag. This is not surprising, as low reddening values ($E(B-V)<0.05$ mag) are
ubiquitous for stars within 90 pc in the general direction of HD 95086
\citep{Reis11}. Based on this discussion, we adopt $A_V$ =
0.02\,$\pm$\,0.02 mag as a reasonable estimate that brackets the range
of plausible extinctions and confirm the spectral type discussion in \citet{Rameau13b}.

We calculate an absolute magnitude of $M_V$ $\simeq$ 2.6, which places
the star near the main sequence (MS). The combined constraints of spectral type A8\,$\pm$1
(assumed uncertainty) and restricting the plausible reddening value to
be $E(B-V)<0.05$ leads to an effective temperature estimate of
$7550\pm100$ K (log($T_{\text{eff}}$) = 3.878\,$\pm$\,0.006) and \textit{V}-band
bolometric correction of BC$_V$ = 0.040\,$\pm$\,0.002 mag
(statistical) $\pm$\,0.026 mag (systematic), where the systematic
uncertainty reflects differences in the bolometric magnitude scales
among six studies \citep{Code76, Balona94, Flower96, Bessell98,
Bertone04, Masana06}. Combining the \textit{Hipparcos} \textit{V} magnitude (assuming
$\pm$\,0.01 mag uncertainty), the revised \textit{Hipparcos} parallax, and our
stated extinction and bolometric magnitude values, we estimate the
following parameters for HD 95086: apparent bolometric flux $f_{\text{bol}}$
= 28.52\,$\pm$\,0.9 pW m$^{-2}$, absolute \textit{V} magnitude $M_V$ =
7.38\,$\pm$\,0.03, absolute bolometric magnitude 2.60\,$\pm$\,0.09,
log($L/L_{\odot}$) = 0.863\,$\pm$\,0.035 dex.

\begin{figure}
 \centering
  \epsscale{1.0}
 \plotone{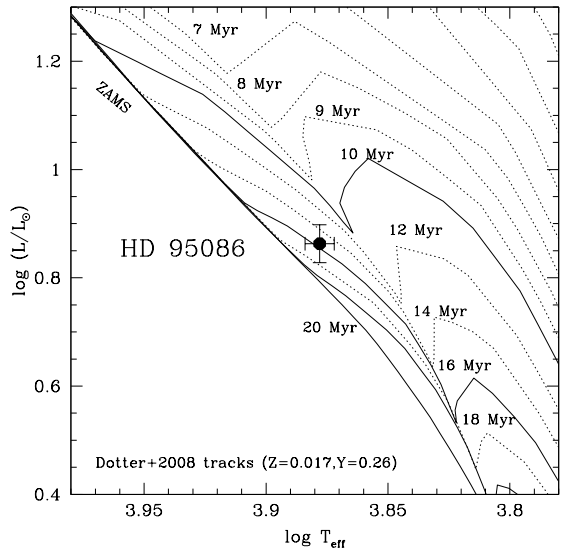}
 \caption{H-R diagram for HD 95086 with Dartmouth isochrones overlaid.}
 \label{HRD}
\end{figure}

In \autoref{HRD},we plot the H-R diagram point for HD 95086 with
Dartmouth isochrones from \citet{Dotter08} adopting approximately
protosolar composition of \textit{Y} = 0.26 and \textit{Z} = 0.017. The star is slightly
above the zero-age main sequence (ZAMS), and appears consistent with pre-MS
age 16\,$\pm$\,2 Myr. Using three other sets of pre-MS evolutionary
tracks, we find the following ages: 15 Myr \citep{D'Antona97}, 17 Myr
\citep{Yi03}, and 24 Myr \citep{Siess97}. Note that models are in
general agreement on the mass of HD 95086: $\sim$1.7 $M_{\odot}$.
Taking into account the different models and observational
uncertainties, we estimate a final individual pre-MS isochronal age
estimate of HD 95086 of 17 Myr ($\pm$2 Myr statistical, $\pm$4 Myr
systematic, total uncertainty $\sim\,\pm$4 Myr). This is
commensurate with the mean LCC subgroup age \citep{Mamajek02, Pecaut12}.

Two factors which could conspire to ruin our age estimate are (1)
unresolved binarity (which would move the star older, closer to the
ZAMS), and (2) if the star is an interloper whose H-R diagram position
and kinematics could conspire to masquerade the star appear as a
likely LCC member. In the unlikely scenario that the star is an
interloper, its H-R diagram position would be consistent with an MS isochronal age of $\sim$270 Myr.

\subsection{Color Constraints}
\label{sec:color_constraint}

At the angular separation of the planet ($0\farcs624$), we are sensitive 
to an \textit{H}-band contrast of 12.7 mag relative to the host star for 
a 5$\sigma$ point source detection. This is an apparent magnitude of 
19.57 mag. Using the \citet{Rameau13b} apparent \textit{L'} magnitude of 16.49, we derive an 
$H-L'$ lower limit of 3.1$\pm$0.5 mag for the planet at 5$\sigma$. 

The expected \textit{H}-band contrast of the planet, given the measured \textit{L'} magnitude \citep{Rameau13b}, 
is 12.03$\pm$1.0 mag based on the COND model \citep{Baraffe03} and 12.50$\pm$1.0 mag 
based on the BT-SETTL model (\citealt{Allard13}, ; \autoref{contrast_curve}). 
The error is propagated from the uncertainty in the \textit{L'} magnitude in \citet{Rameau13b}. 
The contrast curve demonstrates that our data are 
sensitive enough to detect a planet for the COND models (S/N$>\sim$4) and the BT-SETTL models (S/N$>\sim$2.5).
Instead, we detect no point sources at such brightness in our \textit{H}-band data. 

The expected \textit{H} magnitudes for an unreddened A/F star, M dwarf, and T dwarf interlopers which are at the same distance as HD 95086 
are 16.5, 17.5, and 18.5 (see triangle, \autoref{contrast_curve}), 
respectively. This is far brighter than our 19.57 detection limit at the angular separation of the companion. 
If the companion were a T dwarf at the distance of HD 95086 or closer, we would have seen it based on our color and sensitivity 
limits. An unreddened K giant would have to be at 40 kpc in order to have the same \textit{L'} magnitude as the companion, 
and its $H-L'$ color would be 0.17, which we would have detected. The companion is very unlikely to 
be a background source due to the proper motion seen in \citet{Rameau13b}. 

\begin{figure}
 \centering
  \epsscale{1.1}
 \plotone{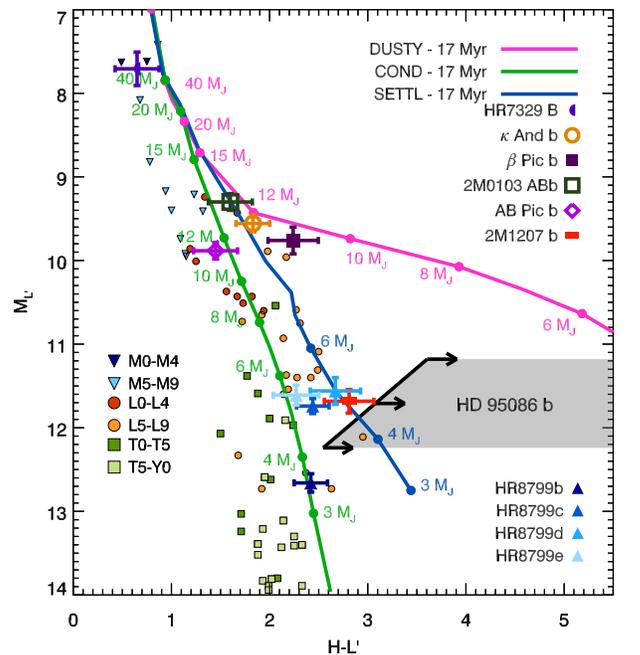}
 \caption{Color magnitude diagram showing the $M_{L'}$ vs. $H-L'$ color 
 of various substellar objects as well as tracks from evolutionary and atmospheric models. 
 The shaded area is defined as a lower limit on the color of the HD 95086 companion.(A color 
 version of this figure is available in the online journal.)}
 \label{cmd}
\end{figure}

We compared the constraints on the photometry of HD 95086 b (shaded area) in a color--magnitude diagram (\autoref{cmd}) to the 
photometry of field M, L, and T dwarfs \citep{Leggett13}, of young (12--30 Myr) companions (HR7329 B, 
\citet{Neuhauser11}; Kappa And b, \citet{Carson13,Bonnefoy13b}; $\beta$ Pic b, \citet{Lagrange10}, 
\citet{Bonnefoy13}; 2M0103 ABb, \citet{Delorme13}; AB Pic b, \citet{Chauvin05}; 2M1207 b, \citet{Chauvin04}; 
HR8799 bcde, \citet{Marois08}, \citet{Skemer12}, and to the COND model, 
the DUSTY model \citet{Chabrier00}, and the BT-SETTL model generated for an age of 17 Myr. 
The $H-L'$ color of HD 95086 b makes it one of the reddest companions directly imaged so far. 
It is inconsistent with COND models, consistent with 4--5 $M_{\text{Jup}}$ BT-SETTL models, 
and consistent DUSTY models at an age of 17 Myr. The color is also at least 2 mag redder than 
the ones of typical M field dwarfs and 1 mag redder than the colors of early-L dwarfs. The location of the 
companion with respect to the sequence of field brown dwarfs and to evolutionary tracks suggest a high dust content in its 
photosphere (possibly under the form of thick clouds; \citealt{Currie11}). In summary, current constraints 
on the photometry of HD 95086 b further suggest that the source is likely a bound companion with peculiar atmospheric 
properties related to low surface gravity. A new astrometric epoch is still mandatory to confirm that the 
companion is comoving with the star, and then exclude the possibility that it could be an extincted or 
intrinsically red background or foreground object.

\subsection{Proper Motion of Background Sources}
There are 20 additional point sources visible in the \textit{H}-band distortion corrected images, ranging from 2$\arcsec$ to 10$\arcsec$ away from 
HD 95086. Using NICMOS \textit{HST} data from 2007
(PI: J. Rhee), we examine the proper motion of these sources relative to HD 95086 with a five-year 
baseline. The proper motion of all the point sources in the field of view are consistent with background sources.

\section{Conclusion}

We analyze \textit{H}-band NICI data aiming to re-detect the planet HD 95086 b. We detect 
no point source in our data, verified through three independent pipelines. 
Our deep dataset provides an $H-L'$ lower limit, 
which rules out foreground L/T dwarfs and distant background K giant contaminants. If the 
object is bound to HD 95086, our color limit is inconsistent with the COND model, 
marginally consistent with the BT-SETTL model, and consistent with the DUSTY model 
at our derived age of 17$\pm$4 Myr.
Future astrometric observations are necessary to establish the nature of this object.
This extremely red object demonstrates the importance of \textit{L'} band for planet detection as 
planets may be much redder than models predict.

\acknowledgments
{\bf Acknowledgements}: We thank the anonymous referee for their comments which improved this paper.
V. Bailey is supported by the National Science Foundation Graduate Research Fellowship (NSF DGE-1143953).
J. Rameau, G. Chauvin, and A.-M. Lagrange acknowledge financial support from the French National Research Agency 
(ANR) through project grant ANR10-BLANC0504-01. 
This letter makes use of the Vizier Online Data Catalog.

\bibliographystyle{apj}

\end{document}